\documentclass[11pt,letterpaper]{article}
\usepackage[T1]{fontenc}
\usepackage[utf8]{inputenc}
\usepackage{lmodern}
\usepackage[margin=1in,headheight=14pt]{geometry}
\usepackage{amsmath,amssymb,amsthm,mathtools}
\usepackage{microtype,booktabs,array,multirow}
\usepackage{enumitem,needspace}
\usepackage{fancyhdr}
\usepackage[hidelinks]{hyperref}
\usepackage[nameinlink,noabbrev,capitalise]{cleveref}
\hypersetup{pdftitle={Mixing Time of the Random-Pair Monte Shuffle},pdfsubject={A manuscript on public random transpositions with private fair swap bits}}
\numberwithin{equation}{section}
\newtheorem{theorem}{Theorem}[section]
\newtheorem{lemma}[theorem]{Lemma}
\newtheorem{proposition}[theorem]{Proposition}
\newtheorem{corollary}[theorem]{Corollary}
\theoremstyle{definition}

\theoremstyle{remark}

\newcommand{\E}{\mathbb E}
\newcommand{\Pp}{\mathbb P}
\newcommand{\R}{\mathbb R}
\newcommand{\Sn}{S_n}
\newcommand{\U}{U_n}
\newcommand{\Id}{\ensuremath{\mathsf{id}}}
\newcommand{\one}{\mathbf 1}
\newcommand{\TV}{\mathrm{TV}}
\newcommand{\Frob}{\mathrm{F}}
\newcommand{\tr}{\operatorname{tr}}
\newcommand{\Tr}{\operatorname{Tr}}
\newcommand{\Var}{\operatorname{Var}}
\newcommand{\Law}{\mathcal L}
\DeclareMathAlphabet{\mathTimesRoman}{OT1}{ptm}{m}{n}
\newcommand{\Tcal}{\mathTimesRoman{T}}
\newcommand{\Chi}{\mathrm{X}}
\newcommand{\Real}{\mathsf R}
\newcommand{\Ideal}{\mathsf I}
\newcommand{\norm}[1]{\left\lVert #1\right\rVert}
\newcommand{\inner}[2]{\left\langle #1,#2\right\rangle}
\newcommand{\eps}{\varepsilon}
\newcommand{\ceil}[1]{\left\lceil #1\right\rceil}

\newcommand{\pos}[1]{\left[#1\right]_+}
\newcommand{\sample}{\xleftarrow{\$}}

\title{\bfseries Mixing Time of the Random-Pair Monte Shuffle}
\author{Ilya Mironov}
\date{\small\today}

\begin{document}
\maketitle
\vspace{-1.2em}
\begin{abstract}
In the random-pair Monte shuffle on $n$ cards, each step chooses two positions independently and uniformly and exchanges their cards with probability $1/2$. The chosen positions are public; the decisions to exchange cards remain hidden. We prove that the mixing time in expected conditional total variation distance is $2n\log n+O(n)$, with an exact leading constant and an $O(n)$ cutoff window. To our knowledge, this is the first public-schedule Monte shuffle whose full-permutation mixing time has been determined up to an additive $O(n)$ term.

The upper bound follows by comparing two executions with the same public schedule to a classical random-transposition walk. For the lower bound, we construct a schedule-dependent test function and estimate its variance using a matrix-averaging operator. Combined with the depth bound of Czumaj and Lingas, the mixing estimate gives public switching networks with at most $2n\log n+O(n)$ switches and $O(\log n)$ depth for fixed error and failure probabilities.

The author obtained this result through interaction with GPT-6 Astra.
\end{abstract}

\section{Introduction}
\label{sec:intro}

The random-transposition shuffle chooses two positions independently and uniformly at each step and exchanges the cards at those positions. Diaconis and Shahshahani~\cite{DS81} proved that its mixing time is $\frac12 n\log n+O(n)$. We ask how the mixing time changes when the chosen positions are revealed and the only hidden choice is whether to exchange the two cards.

We study the \emph{random-pair Monte shuffle}, which exchanges the cards at a uniformly sampled pair of positions with probability $1/2$. The sequence of pairs, called the \emph{schedule}, is public, and the exchange decisions are independent hidden bits. For each schedule we measure the total variation distance of the resulting distribution on permutations from uniform, and then average this distance over schedules.

In this paper we prove that the mixing time is $2n\log n+O(n)$.  Revealing the schedule therefore doubles the leading term relative to the ordinary $1/2$-lazy random-transposition shuffle, whose mixing time is $n\log n+O_\varepsilon(n)$. Together with the factor of two due to laziness, this gives a factor of four relative to classical random transpositions.

Our shuffle is an instance of the Monte framework introduced by Morris~\cite{Morris09}. Here we consider shuffles specified by a public schedule of \emph{fair switches}, each of which either exchanges the cards at two specified positions or leaves them in place, with equal probabilities. The schedule may be fixed, as in the Thorp shuffle, or random, as in our model. Several Monte shuffles are known to mix in $\Theta(n\log n)$ individual switches, including the Thorp shuffle by a recent result~\cite{OpenAIThorp26}.

Our bounds determine both the leading constant and an $O(n)$ cutoff window. Saloff-Coste and Zúñiga, in their refined analysis of classical random transpositions, observe that ``It is quite rare to be able to capture the mixing time of a chain with such precision''~\cite{SCZ08}. To our knowledge, the random-pair Monte shuffle is the first public-schedule Monte shuffle whose full-permutation mixing time has been determined up to an additive $O(n)$ term.

The public schedule defines a switching network, with the hidden bits determining the switch settings. Such networks have cryptographic applications in constructions of Oblivious RAM~\cite{GO96} and format-preserving encryption~\cite{HMR12}. For an oblivious shuffle based on pairwise exchanges, a switching network captures its threat model: the adversary learns the network from the memory access pattern, while the switch settings remain hidden. Each step reads two records, re-encrypts them in private memory, and writes back, concealing whether they were exchanged. Security of shuffling-based Oblivious RAM requires the final permutation to remain indistinguishable from uniform even given the observed access pattern. Format-preserving encryption can be constructed from a switching network as long as the trajectory of any card can be efficiently traced in both directions.

Disjoint switches commute and can be executed in parallel, provided that their dependencies are respected. Combining our mixing bound with the dependency graph depth estimate of Czumaj and Lingas~\cite{CL23} gives, for fixed error and failure probabilities, public switching networks with at most $2n\log n+O(n)$ switches and $O(\log n)$ depth. The probability is over the choice of the network; the mixing guarantee is conditional on that choice. Explicit bounds are given in Section~\ref{sec:depth}.

\subsection{Notation}
\label{sec:notation}
All logarithms are natural unless a base is indicated. We use the following notation:
\begin{center}
\begingroup
\small
\setlength{\tabcolsep}{3pt}
\renewcommand{\arraystretch}{1.05}
\begin{tabular}{@{}>{$}l<{$}@{\quad}p{0.78\linewidth}@{}}
n & Number of cards.\\
{[n]} & $\{1,\ldots,n\}$.\\
\Sn & Symmetric group on $[n]$.\\
\sample & Uniform sampling.\\
\U & Uniform distribution on $\Sn$.\\
\Id & Identity permutation.\\
\one & All-ones vector in $\R^n$.\\
\mathbf e_i & $i$th standard basis vector of $\R^n$.\\
I & Identity matrix.\\
R_\sigma & Permutation matrix of $\sigma\in\Sn$ defined as $R_\sigma\mathbf e_i=\mathbf e_{\sigma(i)}$.\\
\Law(X) & Distribution of a random variable $X$.\\
\oplus & XOR, applied componentwise to bit vectors.\\
\mu*\nu & Convolution: $(\mu*\nu)(\sigma)=\sum_\gamma\mu(\gamma)\nu(\gamma^{-1}\sigma)$.\\
\norm{\mu-\nu}_{\TV} & $\frac12\sum_x|\mu(x)-\nu(x)|$.\\
\chi^2(\mu\|\U) & $n!\sum_{\sigma\in\Sn}\mu(\sigma)^2-1$.\\
d_\infty(\mu,\U) & Relative $L^\infty$ error: $\max_{\sigma\in\Sn}|n!\mu(\sigma)-1|$.\\
\tr(A) & Matrix trace, $\sum_i A_{ii}$.\\
\Tr_W(L) & Operator trace of $L\colon W\to W$.\\
\norm{A}_{\Frob}^2 & Squared Frobenius norm, $\tr(A^{\mathsf T}A)=\sum_{i,j}A_{ij}^2$.\\
\end{tabular}
\endgroup
\end{center}
Permutations map card labels to positions, and products act from right to left.

\Needspace{14\baselineskip}
\subsection{Random-pair Monte shuffle and main results}
\label{sec:model}

We define the \emph{random-pair Monte shuffle} on $n$ cards as follows. At each step $t$, sample $x_t,y_t$ independently and uniformly with replacement from $[n]$, and set $\tau_t=(x_t\,y_t)$, with $\tau_t=\Id$ when $x_t=y_t$. All endpoint samples are independent across steps. Each $\tau_t$ specifies a \emph{switch}.

Independently, sample a private bit vector $B=(b_1,\ldots,b_m)$ uniformly from $\{0,1\}^m$, and set
\begin{equation}
\begin{aligned}
 \text{public schedule:}\quad
 &\Tcal_t=(\tau_1,\ldots,\tau_t),\qquad \Tcal=\Tcal_m,\\[2pt]
 \text{realized permutation:}\quad
 &\pi_t(\Tcal,B)=\tau_t^{b_t}\cdots\tau_1^{b_1},\\[2pt]
 \text{conditional distribution:}\quad
 &\Pi_{\Tcal}=\Law(\pi_m(\Tcal,B)\mid\Tcal).
\end{aligned}
 \label{eq:model}
\end{equation}
For fixed $\Tcal$, the distribution $\Pi_{\Tcal}$ averages over the private bit vector $B$. The function $\pi_t(\Tcal,B)$ uses only the first $t$ entries of each argument; we abbreviate it to $\pi_t$ when the inputs are understood. We take $\Tcal_0$ to be the empty schedule and $\pi_0(\Tcal,B)=\Id$. All $m$ sampled pairs, including equal-endpoint null steps, count toward mixing time. 
Define the conditional error and its schedule average by
\begin{equation}
 \Delta(\Tcal)=\norm{\Pi_{\Tcal}-\U}_{\TV},
 \qquad
 d_n(m)=\E_{\Tcal}\Delta(\Tcal).
 \label{eq:distance}
\end{equation}
Equivalently,
\begin{equation}
 d_n(m)=
 \norm{\Law(\Tcal,\pi_m(\Tcal,B))-\Law(\Tcal)\otimes\U}_{\TV}.
 \label{eq:joint-distance}
\end{equation}
The target is therefore a uniform full permutation independent of its public schedule. For $0<\eps<1$, let
\[
 m_\eps(n)=\min\{m\ge0:d_n(m)\le\eps\}.
\]

\Needspace{12\baselineskip}
\begin{theorem}[Main result]
\label[theorem]{thm:main}
For every fixed $0<\eps<1$, the mixing time of the random-pair Monte shuffle satisfies
\begin{equation}
 m_\eps(n)=2n\log n+O_\eps(n).
 \label{eq:mixing-time}
\end{equation}
Moreover, for $n\ge10$ and an integer $m=2n\log n+cn\ge0$,
\begin{align}
 c<0&\quad\Longrightarrow\quad d_n(m)\ge1-8e^{c/2},
 \label{eq:left-window}\\
 c>0&\quad\Longrightarrow\quad d_n(m)\le8e^{-c/2}.
 \label{eq:right-window}
\end{align}
\end{theorem}

We prove \cref{thm:main} in~\cref{sec:mixing}. Doubling the leading term in the number of steps suffices for the stronger conditional pointwise guarantee of \cref{prop:conditional-pointwise}. For fixed $0<\eps,\delta<1$, a schedule of length $4n\log n+O_{\eps,\delta}(n)$ satisfies
\[
 \frac{1-\eps}{n!}\le\Pi_{\Tcal}(\sigma)\le\frac{1+\eps}{n!}
 \qquad\text{for every }\sigma\in\Sn
\]
with probability at least $1-\delta$ over the public schedule.

For the upper bound in~\cref{eq:right-window}, we run two copies of the shuffle with the same schedule and independent private bits. After averaging over schedules, their relative permutation follows the ordinary $1/2$-lazy random-transposition walk. The collision probability of the two copies is therefore the return probability of this auxiliary walk. Its normalized excess over $1/n!$ equals the expected conditional chi-square error of our shuffle. We bound this error using the $L^2$ estimate of Saloff-Coste and Z\'u\~niga~\cite{SCZ08}, then apply the standard inequality relating total variation and chi-square distance. Ordinary TV mixing of the auxiliary walk is not sufficient: the argument needs control of its return probability. This accounts for the additional factor of two beyond laziness.

For the lower bound (\cref{eq:left-window}), we construct a test function of the schedule and the observed permutation. Its value is the expected number of cardwise agreements with an independent run under the same schedule, centered by subtracting one. This function aggregates the small biases in the destinations of individual cards. To turn the difference in means into a distinguisher, we control the statistic’s variance under the shuffle and uniform distributions. A change of schedule expresses its second moment as the trace of a matrix-averaging operator. We decompose the matrix space into four subspaces preserved by relabeling and compute the eigenvalue on each. The resulting variance estimate, together with Chebyshev's inequality, gives the matching lower bound.

\subsection{Related work}
\label{sec:related}

Morris~\cite[Section~3.1]{Morris09} introduced the collision/Monte framework and developed entropy methods for shuffles built from fair identity/transposition choices. A \emph{fair switch} on positions $x,y$ is
\[
 c^\$(x,y)=(x\,y)^b,\qquad b\sample\{0,1\}.
\]
Each occurrence uses a fresh, independent private bit. A \emph{step} applies one candidate switch; a \emph{round} applies a complete layer of disjoint switches.

We review Monte shuffles and random-transposition shuffles below. Table~\ref{tab:shuffles} summarizes their definitions and mixing bounds.

\paragraph{The Thorp shuffle.}
For $n=2^\ell$, identify positions with $\mathbb F_2^\ell$ and let $\mathbf e_0,\ldots,\mathbf e_{\ell-1}$ be its standard basis. In round $i$, independently apply a fair switch to every unordered pair $\{x,x\oplus\mathbf e_{i\bmod\ell}\}$. This is the Thorp shuffle in a rotating coordinate frame. In the equivalent time-invariant description, the deck is cut in half and corresponding cards from the two halves are interleaved in independently chosen random orders~\cite{Thorp73}. The first polylogarithmic bound on the Thorp shuffle mixing time was Morris's $O(\log^{44}n)$ estimate on the number of rounds~\cite{Morris08}, which was subsequently improved to the asymptotically optimal $\Theta(\log n)$~\cite{MT06,Morris09,Morris13,jiang2026subquadraticupperboundmixing,OpenAIThorp26}.

\paragraph{XOR Swap-Or-Not.}
Hoang, Morris, and Rogaway~\cite{HMR12} introduced the Swap-Or-Not shuffle. In its XOR form, round $i$ samples a public key $K_i\in\mathbb F_2^\ell$ and applies independent fair switches to all pairs $\{x,x\oplus K_i\}$. Conditional on the round keys, their proof bounds convergence in $O(\log n)$ rounds for the joint positions of any prescribed fixed fraction of the cards, not a total variation bound for the full permutation. Dai, Hoang, and Tessaro~\cite{DHT17} obtain tighter Swap-Or-Not bounds
using the chi-squared method. Jho and Lee~\cite{JL23} generalize Swap-Or-Not by replacing the XOR pairings with keyed partitions, reducing the round count for the same fractional-deck guarantee.

Ristenpart and Yilek's Mix-and-Cut shuffle~\cite{RY13} obtains full-permutation statistical security by recursively randomizing the partition of a deck into two halves. For $n=2^\ell$, their construction applies $r$ rounds of Swap-Or-Not to each block, cuts it in half, and independently shuffles the two halves by the same procedure. For fixed error, their analysis gives $O_\eps(\log^2 n)$ layers and a total of $O_\eps(n\log^2 n)$ switches.

Morris and Rogaway's Sometimes-Recurse shuffle~\cite{MR14} recurses on only one half after each Swap-Or-Not stage. With independent randomness at each stage, it is a Monte shuffle whose public schedule consists of matching layers on successively smaller subdecks. Their recursive argument, combined with the Swap-Or-Not bound on the partial deck mixing, gives full-permutation conditional TV mixing in $O_\eps(n\log n)$ switches and $O_\eps(\log^2 n)$ layers.

\paragraph{Random matching layers.} 
One can construct a switching network by sampling a uniform perfect matching in each round. Czumaj~\cite{Czumaj15} proves that perfect-matching networks mix any prescribed fixed fraction of the cards in $O(\log n)$ layers with probability $1-O(n^{-2})$ over the choice of the network. The same work gives a separate explicit full-permutation network with $O(n\log n)$ switches and $O(\log^2 n)$ depth. 

\paragraph{The iid random-pair upper bound.}
In the work closest to ours, Anari~\cite{Anari26} proves multiplicative pointwise near-uniformity for a random-pair Monte shuffle, with high probability over the public schedule, in $O(n\log n)$ steps. We obtain the sharp leading constant for conditional TV mixing and a matching lower bound. Our multiplicative pointwise guarantee uses $4n\log n+O(n)$ steps, twice the leading count for conditional total variation.

Czumaj and Lingas~\cite{CL23} show by a counting argument that $m=\Omega(n\log n)$ independent uniform distinct-pair interactions have dependency depth $\Theta(m/n)$ in expectation and with high probability. We apply this result in Section~\ref{sec:depth}.

\paragraph{Random transpositions.}
The random transposition shuffle samples $x,y$ independently and uniformly from $[n]$ and applies $(x\,y)$.  Diaconis and Shahshahani~\cite{DS81} proved the sharp total-variation upper bound of $\frac12 n\log n+O_\eps(n)$ steps for each fixed error $0<\eps<1$. Saloff-Coste and Z\'u\~niga~\cite{SCZ08} refined the classical $L^2$ estimates with explicit constants in discrete and continuous time; we use their continuous-time bound in spectral form (\cref{prop:upper}).

A matching lower bound follows by tracking the number of untouched cards, known as the coupon-collector problem. In a series of works, this lower bound was shown to be optimal in several senses: by White for convergence in separation distance~\cite{White19}, by Teyssier through an exact total-variation cutoff profile~\cite{Teyssier20}, and by Jain and Sawhney~\cite{JS26}, who proved that the walk is asymptotically uniform at the first time every card has been touched.

\paragraph{Semi-random transpositions.}
Fix a sequence of positions $x_1,x_2,\ldots$ before execution. At step~$i$, sample $y_i$ uniformly from $[n]$ and apply $(x_i\,y_i)$. Mossel, Peres, and Sinclair~\cite{MPS04} proved an $O_\eps(n\log n)$ mixing bound uniformly over the fixed sequence. Saloff-Coste and Z\'u\~niga~\cite{SCZ07} and, independently, Ganapathy~\cite{Gan07} sharpened this to $n\log n+O_\eps(n)$, with the leading constant optimal over all fixed endpoint sequences, as witnessed by the star-transposition shuffle.

\begin{table}[tp]
\centering
\caption{Shuffling schemes on $S_n$ and their mixing guarantees conditional on public information (boxed variables). Bounds are for fixed total variation distance from the uniform distribution $0<\eps<1$; $\dagger$ marks bounds on the joint positions of any prescribed $\lfloor(1-\alpha)n\rfloor$ cards, with fixed $\alpha\in(0,1)$; $\ddagger$ is convergence in the multiplicative point-wise distance. A step is one transposition or a fair switch; a round is one layer of disjoint switches. Each $c^\$(x,y)=(x\,y)^b$ uses a fresh private fair bit $b$. The first three schemes assume $n=2^\ell$; each deck is identified with $\mathbb F_2^{\ell}$, and the $\mathbf e_j$ are standard basis vectors. References are not exhaustive.}
\label{tab:shuffles}
\begingroup
\small
\setlength{\tabcolsep}{4pt}
\renewcommand{\arraystretch}{1.3}
\begin{tabular}{@{}>{\raggedright\arraybackslash}m{100pt}
                    >{\raggedright\arraybackslash}m{179pt}
                    >{\raggedright\arraybackslash}m{110pt}
                    >{\raggedright\arraybackslash}m{56pt}@{}}
\toprule
\textbf{Name} & \textbf{Definition} & \textbf{Mixing time} & \textbf{References} \\
\midrule
\multirow{2}{=}{Thorp shuffle}
& \multirow{2}{=}{$T_i=\{\{x,y\}\colon x\oplus y=\mathbf e_{i\bmod\ell}\}$\newline
  apply $\prod_{\{x,y\}\in T_i}c^\$(x,y)$}
& $O_\eps(\log^3 n)$ rounds & \cite{Morris13} \\
\cmidrule(lr){3-4}
& & $\Theta_\eps(\log n)$ rounds & \cite{OpenAIThorp26} \\
\midrule
Swap-or-Not
& $\boxed{K_i}\sample\mathbb F_2^\ell$\newline
  $T_i=\{\{x,y\}\colon x\oplus y=K_i\}$\newline
  apply $\prod_{\{x,y\}\in T_i}c^\$(x,y)$
& $O_{\eps,\alpha}(\log n)^\dagger$ rounds
& \cite{HMR12} \\
\midrule
Sometimes-Recurse
& For $i=\ell,\dots,1$, run $r$ rounds of
  Swap-or-Not on the first $2^i$ positions
& $r=O_\eps(\log n)$\newline
$O_\eps(n\log n)$ steps total
& \cite{MR14} \\
\midrule
Random perfect-matching Monte
& $T_i=\boxed{\{\{x_j,y_j\}\}_{j=1}^{n/2}}$ uniform\newline
  random partition of $[n]$\newline
apply $\prod_{\{x,y\}\in T_i} c^\$(x,y)$
& $O_{\eps,\alpha}(\log n)^\dagger$ rounds
& \cite{Czumaj15} \\
\midrule
\multirow{2}{=}{Random-pair Monte}
& \multirow{2}{=}{$\boxed{x,y}\sample[n]$\newline
  apply $c^\$(x,y)$}
& $O_\eps(n\log n)^\ddagger$ steps & \cite{Anari26} \\
\cmidrule(lr){3-4}
& & $2n\log n+O_\eps(n)$ steps & This work \\
\midrule
Semi-random transpositions
& $\boxed{x_i}\in[n]$ fixed in advance; $y_i\sample[n]$\newline
apply $(x_i\,y_i)$
& $O_\eps(n\log n)$ steps
& \cite{MPS04} \\
\midrule
Random transposition
& $x,y\sample[n]$\newline
  apply $(x\,y)$
& $\tfrac12n\log n+O_\eps(n)$ steps
& \cite{DS81} \\
\bottomrule
\end{tabular}
\endgroup
\end{table}

\section{Upper and matching lower bounds for conditional mixing}
\label{sec:mixing}

We first prove the upper bound by comparing two executions with a common schedule. We then construct a test function for the lower bound and estimate its first two moments.

\subsection{The shared-schedule collision identity}
\label{sec:collision}

Let $P$ be the classical random-transposition kernel: independently sample $x,y$
uniformly from $[n]$ and apply $(x\,y)$, including the identity when $x=y$.
Our schedule-averaged kernel is its $1/2$-lazification,
\[
 Q=\frac{I+P}{2}.
\]
Write $q_m=\Law(\pi_m(\Tcal,B))=\E_{\Tcal}\Pi_{\Tcal}$ for its
$m$-step distribution from $\Id$. Define the expected conditional
chi-square error and the pointwise relative error of the
\emph{schedule-averaged} distribution by
\[
 \Chi_n(m)=\E_{\Tcal}\chi^2(\Pi_{\Tcal}\|\U),\qquad
 \eta_m=d_\infty(q_m,\U)
       =\max_{\sigma\in\Sn}|n!q_m(\sigma)-1|.
\]
Note that $\eta_m$ concerns the averaged distribution $q_m$.
The conditional relative error $d_\infty(\Pi_{\Tcal},\U)$ is a different
quantity, which we bound in \cref{prop:conditional-pointwise}.

\Needspace{9\baselineskip}
\begin{lemma}[Shared-schedule collision identity]
\label[lemma]{lem:collision}
For every $n\ge2$ and integer $m\ge0$,
\begin{equation}
 \Chi_n(m)=n!q_m(\Id)-1=\eta_m.
 \label{eq:collision}
\end{equation}
\end{lemma}

\begin{proof}
Run two copies $X_t=\pi_t(\Tcal,B)$ and
$Y_t=\pi_t(\Tcal,B')$ with the same public schedule and independent
private bits. Their relative permutation $Z_t=X_t^{-1}Y_t$ evolves as
\[
 Z_t=(X_{t-1}^{-1}\tau_t X_{t-1})^{b_t\oplus b'_t}Z_{t-1}.
\]
Conditional on the past, the conjugated endpoints are still
independent and uniform, and $b_t\oplus b'_t$ is a fresh fair bit.
Hence $Z_t$ has transition kernel $Q$ and starts at $\Id$.
Conditional on $\Tcal$, the walks $X_m$ and $Y_m$ are independent draws
from $\Pi_{\Tcal}$. Thus, for every $\sigma\in\Sn$,
\[
 q_m(\sigma)
 =\E_{\Tcal}\sum_{\pi\in\Sn}\Pi_{\Tcal}(\pi)\Pi_{\Tcal}(\pi\sigma).
\]
At $\sigma=\Id$, this gives $n!q_m(\Id)-1=\Chi_n(m)$.
Subtracting the uniform baseline, we have
\[
 q_m(\sigma)-\frac1{n!}
 =\E_{\Tcal}\sum_{\pi\in\Sn}
   \left(\Pi_{\Tcal}(\pi)-\frac1{n!}\right)
   \left(\Pi_{\Tcal}(\pi\sigma)-\frac1{n!}\right).
\]
The map $\pi\mapsto\pi\sigma$ permutes $\Sn$, so the two
vectors in this inner product have the same Euclidean norm.
Cauchy--Schwarz for each fixed schedule, followed by averaging, gives
\[
 |n!q_m(\sigma)-1|
 \le n!\,\E_{\Tcal}\sum_{\pi\in\Sn}
       \left(\Pi_{\Tcal}(\pi)-\frac1{n!}\right)^2
 =\Chi_n(m).
\]
Taking the maximum over $\sigma$, with equality at $\Id$, proves
\eqref{eq:collision}.
\end{proof}

The collision identity explains the additional factor of two relative
to the ordinary lazy random-transposition shuffle. Since $Q$ is symmetric,
the ordinary chi-square error satisfies
\[
 \chi^2(q_m\|\U)=n!q_{2m}(\Id)-1,
\]
whereas the expected conditional chi-square error is
$\Chi_n(m)=n!q_m(\Id)-1$ by \eqref{eq:collision}.
Thus a chi-square bound for the ordinary lazy walk gives the same bound
on the expected conditional error at twice the time. This explains the
additional factor of two in our upper bound, beyond the slowdown from laziness.
The matching lower bound shows that this additional factor is also
necessary for conditional TV mixing. The next subsection obtains the
required return-probability bound
from the $L^2$ estimate of Saloff-Coste and Z\'u\~niga~\cite{SCZ08}.

\subsection{Conditional mixing upper bound}

Let $p_0=1,p_1,\ldots,p_{n!-1}$ be the eigenvalues of $P$,
counted with multiplicity, with $p_0$ corresponding to constant
functions. Proposition~4.6 of Saloff-Coste and Z\'u\~niga~\cite{SCZ08}
gives the following estimate in our normalization:
\begin{equation}
 n\ge10,\quad c\ge4
 \quad\Longrightarrow\quad
 \sum_{j\ne0}
 \exp\!\left[-n\left(\log n+\frac c2\right)(1-p_j)\right]
 \le e^{4-c}.
 \label{eq:classical}
\end{equation}
We use their continuous-time $L^2$ bound only in this spectral form.

\begin{proposition}[Averaged pointwise and conditional-TV upper bounds]
\label[proposition]{prop:upper}
For $n\ge10$ and $c\ge4$,
\begin{equation}
 m\ge2n\log n+cn
 \quad\Longrightarrow\quad
 \Chi_n(m)=\eta_m\le e^{4-c},
 \qquad d_n(m)\le\tfrac12 e^{2-c/2}.
 \label{eq:upper}
\end{equation}
\end{proposition}

\begin{proof}
By \eqref{eq:collision} and translation invariance,
$n!q_m(\Id)=\tr(Q^m)$. Since $Q=(I+P)/2$,
\begin{align*}
 \eta_m=\Chi_n(m)
 &=\tr(Q^m)-1
 =\sum_{j\ne0}\left(\frac{1+p_j}{2}\right)^m\\
 &\le\sum_{j\ne0}\exp\!\left[-\frac m2(1-p_j)\right]
 \le e^{4-c}.
\end{align*}
The first inequality uses $z^m\le e^{-m(1-z)}$ for
$0\le z\le1$; the last uses $m/2\ge n\log n+cn/2$ and
\eqref{eq:classical}.

Finally, Cauchy--Schwarz for each conditional distribution
and Jensen's inequality give
\begin{equation}
 d_n(m)
 \le\frac12\E_{\Tcal}\sqrt{\chi^2(\Pi_{\Tcal}\|\U)}
 \le\frac12\sqrt{\Chi_n(m)}
 \le\frac12 e^{2-c/2}.
 \label{eq:tv-upper-second}
\end{equation}
\end{proof}

For $n\ge10$ and $0<\eps\le1/2$, a sufficient number of sampled steps for
joint error at most $\eps$ is
\begin{equation}
 m\ge
 \ceil{2n\left(\log n+2+\log\frac1{2\eps}\right)}.
 \label{eq:eps-upper}
\end{equation}
We can also bound the probability of a bad schedule. Since $\Delta(\Tcal)^2\le\frac14\chi^2(\Pi_{\Tcal}\|\U)$,
\begin{equation}
 \Pp_{\Tcal}\{\Delta(\Tcal)>\eps\}
 \le\frac{\Chi_n(m)}{4\eps^2}.
 \label{eq:bad-schedule}
\end{equation}
Consequently, for $n\ge10$, $0<\eps\le1/2$, and $0<\delta<1$, a sufficient
number of sampled steps for this probability to be at most $\delta$ is
\begin{equation}
 m\ge
 \ceil{2n\left(
 \log n+2+\log\frac1{2\eps}+\frac12\log\frac1\delta
 \right)}.
 \label{eq:eps-delta-upper}
\end{equation}

We next obtain a pointwise bound conditional on the schedule.
The argument uses two independent blocks of the shuffle. Write
\[
 \Delta_\infty(\Tcal)=d_\infty(\Pi_{\Tcal},\U)
 =\max_{\sigma\in\Sn}|n!\Pi_{\Tcal}(\sigma)-1|.
\]
This is a schedule-dependent bound, unlike in
$\eta_m=d_\infty(q_m,\U)$.

\begin{proposition}[Conditional relative $L^\infty$ mixing]
\label[proposition]{prop:conditional-pointwise}
Let $n\ge10$, $c\ge4$, and let $m\ge2n\log n+cn$ be an integer.
For an iid public schedule $\Tcal$ of length $2m$,
\begin{equation}
 \E_{\Tcal}\Delta_\infty(\Tcal)\le e^{4-c}.
 \label{eq:conditional-pointwise-mean}
\end{equation}
In particular, for $0<\eps,\delta<1$, a schedule of length
\begin{equation}
 M=2\ceil{2n\log n+n\left(4+\log\frac1{\eps\delta}\right)}
 \label{eq:conditional-pointwise-length}
\end{equation}
satisfies, with probability at least $1-\delta$ over its public schedule,
\begin{equation}
 \frac{1-\eps}{n!}\le\Pi_{\Tcal}(\sigma)\le\frac{1+\eps}{n!}
 \qquad\text{for every }\sigma\in\Sn.
 \label{eq:conditional-pointwise-guarantee}
\end{equation}
\end{proposition}

\begin{proof}
For any two probability distributions $\mu,\nu$ on $\Sn$,
centered convolution and Cauchy--Schwarz give
\[
 d_\infty(\mu*\nu,\U)
 \le\sqrt{\chi^2(\mu\|\U)\,\chi^2(\nu\|\U)}.
\]
Split $\Tcal$ into two independent length-$m$ schedules
$\Tcal^{(1)},\Tcal^{(2)}$. Conditional on these schedules,
$\Pi_{\Tcal}=\Pi_{\Tcal^{(2)}}*\Pi_{\Tcal^{(1)}}$, so
\[
 \E_{\Tcal}\Delta_\infty(\Tcal)
 \le\left(\E_{\Tcal^{(1)}}
       \sqrt{\chi^2(\Pi_{\Tcal^{(1)}}\|\U)}\right)^2
 \le\Chi_n(m)\le e^{4-c}.
\]
Taking $c=4+\log(1/(\eps\delta))$ and applying Markov's
inequality proves \eqref{eq:conditional-pointwise-guarantee}.
\end{proof}

Thus $4n\log n+O_{\eps,\delta}(n)$ steps suffice for fixed
$\eps,\delta$. This is the conditional pointwise guarantee studied
by Anari~\cite{Anari26}, with an explicit
sufficient constant.

\subsection{A schedule-dependent distinguisher}

Counting possible outputs gives the lower bound $2^m\ge(1-\eps)n!$, whose leading term is $(n\log n)/\log2$, which falls short of the goal of proving the tightness of the constant $2$. We instead use a test function that combines the biases in the destinations of all cards, conditional on the schedule.

For a schedule prefix $\Tcal_t$, let
\[
 M_{\Tcal_t}(j,i)=\Pp\{\pi_t(\Tcal,B)(i)=j\mid\Tcal_t\}.
\]
The probability is over $b_1,\ldots,b_t$, with the schedule fixed. Thus column $i$ is the distribution of the position of card $i$. Write $M_{\Tcal}=M_{\Tcal_m}$ for the final matrix. To measure its deviation from uniform, subtract the matrix $J$ whose entries are all $1/n$:
\[
 J(j,i)=\frac1n\quad(i,j\in[n]),\qquad
 C_{\Tcal_t}=M_{\Tcal_t}-J,\qquad
 C_{\Tcal}(j,i)=M_{\Tcal}(j,i)-\frac1n.
\]
The rows and columns of $C_{\Tcal}$ sum to zero. The squared Frobenius norm $\norm{C_{\Tcal}}_{\Frob}^2=\sum_{i,j}C_{\Tcal}(j,i)^2$ measures the total squared single-card destination bias.

Given the schedule and a candidate permutation $\sigma$, consider the test function
\begin{equation}
 F(\Tcal,\sigma)=\sum_{i=1}^nM_{\Tcal}(\sigma(i),i)-1.
 \label{eq:statistic}
\end{equation}
Equivalently, $F(\Tcal,\sigma)=\sum_i C_{\Tcal}(\sigma(i),i)$. The quantity $F(\Tcal,\sigma)+1$ is the expected number of cardwise agreements between~$\sigma$ and an independent execution of the same schedule.

We will distinguish the two distributions by thresholding $F$ in the proof of \cref{thm:lower}.
Let~$\Real$ denote the joint distribution of the schedule and realized permutation, $(\Tcal,\pi_m(\Tcal,B))$, and $\Ideal$ the distribution of $(\Tcal,\sigma)$ with $\sigma\sim\U$ independent of $\Tcal$. By \eqref{eq:joint-distance}, $d_n(m)=\norm{\Real-\Ideal}_{\TV}$.

Throughout the lower-bound proof, assume $n\ge4$ and let
\begin{equation}
 \lambda=1-\frac1n.
 \label{eq:eigenvalues}
\end{equation}

\begin{lemma}[First moments and ideal variance]
\label[lemma]{lem:first}
The statistic \eqref{eq:statistic} satisfies
\begin{equation}
 \E_{\Real}F=(n-1)\lambda^m,
 \qquad
 \E_{\Ideal}F=0,
 \qquad
 \Var_{\Ideal}F=\lambda^m.
 \label{eq:first-moments}
\end{equation}
\end{lemma}

\begin{proof}
For $\tau=(u\,v)$ obtained from independent uniform endpoints $u,v\in[n]$, set
\[
 J_\tau=\frac12(\mathbf e_u-\mathbf e_v)(\mathbf e_u-\mathbf e_v)^{\mathsf T},
 \qquad P_\tau=I-J_\tau.
\]
A fair switch averages the probabilities at its two positions. Thus, if $p$ is a card's position-probability column vector before the switch, it becomes $P_{\tau}p$ afterward: the entries at $u$ and $v$ are both $(p_u+p_v)/2$, and the others are unchanged. When $u=v$, $J_\tau=0$ and $P_\tau=I$.

Since card $i$ starts at position $i$, its initial distribution is $\mathbf e_i$. Applying the updates to all columns gives
\[
 M_{\Tcal_t}=P_{\tau_t}\cdots P_{\tau_1}.
\]
Coordinate averaging fixes $J$ on both sides. Thus the centered matrices satisfy
\[
 C_{\Tcal_0}=I-J,\qquad
 C_{\Tcal_t}=P_{\tau_t}C_{\Tcal_{t-1}},\qquad
 C_{\Tcal}=C_{\Tcal_m}.
\]
Here and below, $\E_\tau$ uses this endpoint-sampling distribution, including $\tau=\Id$ with probability $1/n$. Averaging gives
\[
 \E_\tau J_\tau=\frac{I-J}{n},\qquad
 \E_\tau P_\tau=\lambda I+\frac Jn,\qquad P_\tau^2=P_\tau.
\]
Since $JC_{\Tcal_{t-1}}=0$, symmetry and idempotence of $P_{\tau_t}$ imply
\[
 \E\!\left[\norm{C_{\Tcal_t}}_{\Frob}^2\mid \tau_1,\ldots,\tau_{t-1}\right]
 =\tr\!\left(C_{\Tcal_{t-1}}^{\mathsf T}\E[P_{\tau_t}]C_{\Tcal_{t-1}}\right)
 =\lambda\norm{C_{\Tcal_{t-1}}}_{\Frob}^2.
\]
Starting from $\norm{I-J}_{\Frob}^2=n-1$, iteration yields
\begin{equation}
 \E_{\Tcal}\norm{C_{\Tcal}}_{\Frob}^2=(n-1)\lambda^m.
 \label{eq:matrix-energy}
\end{equation}
Under the actual shuffle, $\Pp\{\pi_m(i)=j\mid\Tcal\}=M_{\Tcal}(j,i)$, with probability over $B\sample\{0,1\}^m$. Hence
\[
 \E_{\Real}[F\mid\Tcal]
 =\sum_{i,j}M_{\Tcal}(j,i)\Pp\{\pi_m(i)=j\mid\Tcal\}-1
 =\sum_{i,j}M_{\Tcal}(j,i)^2-1.
\]
Since $\sum_{i,j}M_{\Tcal}(j,i)=n$, the last expression equals
$\sum_{i,j}(M_{\Tcal}(j,i)-1/n)^2=\norm{C_{\Tcal}}_{\Frob}^2$.
Averaging over $\Tcal$ and applying \eqref{eq:matrix-energy} computes the realized mean.

For an independent uniform permutation, the conditional mean is zero because $C_{\Tcal}$ has zero column sums.
Expanding the square under a uniform permutation gives
\[
 \E_{\sigma\sim\U}\left(\sum_i C_{\Tcal}(\sigma(i),i)\right)^2
 =\frac1{n-1}\norm{C_{\Tcal}}_{\Frob}^2.
\]
Indeed, the diagonal terms contribute $\norm{C_{\Tcal}}_{\Frob}^2/n$.
The remaining terms contribute
$\norm{C_{\Tcal}}_{\Frob}^2/[n(n-1)]$, using the zero row and column sums and the uniform distribution on distinct images. Averaging over~$\Tcal$ and using \eqref{eq:matrix-energy} completes the proof.
\end{proof}

\subsection{The realized second moment}

The mean gap alone is not enough: it could be due to rare large values of $F$. We now estimate the second moment $\E_{\Real}F^2=\E_{\Tcal,B}[F(\Tcal,\pi_m(\Tcal,B))^2]$.  The next lemma absorbs the realized walk into a change of schedule, leaving an expectation over schedules alone.
\Needspace{7\baselineskip}
\begin{lemma}\label[lemma]{lem:three-copy}
With $C_{\Tcal}$ as above,
\begin{equation}
 \E_{\Real}F^2=\E_{\Tcal}\bigl[(\tr C_{\Tcal})^2\bigr].
 \label{eq:three-copy}
\end{equation}
\end{lemma}

\begin{proof}
Let $f(\sigma)=|\{i\colon\sigma(i)=i\}|-1$ be the centered fixed-point count. For a fixed schedule $\Tcal$ and any candidate permutation $\sigma$,
\[
 F(\Tcal,\sigma)=\E_D
   f\bigl(\sigma^{-1}\pi_m(\Tcal,D)\bigr),
\]
where $D$ is uniform on $\{0,1\}^m$.

Fix $\Tcal,B$, and independently sample $\widetilde B$ uniformly from $\{0,1\}^m$. Let
\[
 \pi_t=\pi_t(\Tcal,B),\qquad
 \widetilde\pi_t=\pi_t(\Tcal,\widetilde B).
\]
Both walks use the original schedule $\Tcal$. Define a new schedule by
\[
 \widetilde\tau_t=\pi_{t-1}^{-1}\tau_t\pi_{t-1},\qquad
 \widetilde\Tcal=(\widetilde\tau_1,\ldots,\widetilde\tau_m).
\]
For every choice of $\widetilde B$, the relative walk satisfies
\begin{align*}
 \pi_t^{-1}\widetilde\pi_t
 &=\pi_{t-1}^{-1}\tau_t^{\,b_t\oplus\widetilde b_t}
   \widetilde\pi_{t-1}\\
 &=\widetilde\tau_t^{\,b_t\oplus\widetilde b_t}
   \pi_{t-1}^{-1}\widetilde\pi_{t-1}.
\end{align*}
Iterating from $\pi_0=\widetilde\pi_0=\Id$ yields
\[
 \pi_m^{-1}\widetilde\pi_m
 =\pi_m(\widetilde\Tcal,B\oplus\widetilde B).
\]
The schedule $\widetilde\Tcal$ depends on $\Tcal,B$, but not on $\widetilde B$. Since $B\oplus\widetilde B$ is uniform for fixed $B$, we obtain
\[
 F(\Tcal,\pi_m)
 =\E_{\widetilde B}f(\pi_m^{-1}\widetilde\pi_m)
 =\E_{\widetilde B}f\bigl(\pi_m(\widetilde\Tcal,B\oplus\widetilde B)\bigr)
 =\E_{\widetilde B} f(\pi_m(\widetilde\Tcal,\widetilde B)).
\]
Now fix $B$ and average over $\Tcal$. Conditional on the preceding public transpositions, $\pi_{t-1}$ is fixed and $\tau_t$ is fresh. Conjugation by $\pi_{t-1}^{-1}$ preserves the distribution of its endpoints, including equal endpoints. Thus $\widetilde\Tcal$ has the same distribution as $\Tcal$ for each fixed $B$, and
\[
 \E_{\Real}F^2
 =\E_{\Tcal,B}\left\{\E_{\widetilde B} f(\pi_m(\widetilde\Tcal,\widetilde B))\right\}^2
 =\E_{\Tcal}\left\{\E_B f(\pi_m(\Tcal, B))\right\}^2.
\]
Finally, $\E_B f(\pi_m(\Tcal, B))=\E_{\sigma\sim\Pi_{\Tcal}}f(\sigma)
=\tr M_{\Tcal}-1=\tr C_{\Tcal}$, as claimed.
\end{proof}

Equip $\R^{n\times n}$ with the Frobenius inner product
$\inner{X}{Y}_{\Frob}=\tr(X^{\mathsf T}Y)$.
To evaluate the remaining second moment, consider the matrix space and averaging operator
\[
 \mathcal Z=\{X\in\R^{n\times n}:X\one=X^{\mathsf T}\one=0\},
 \qquad
 \mathcal A(X)=\E_\tau[P_\tau X P_\tau]\quad(X\in\mathcal Z).
\]
Each update averages two rows and the corresponding columns, preserving $\mathcal Z$. By independence of the public pairs, for $X\in\mathcal Z$ we have
\[
 \mathcal A^m(X)
 =\E_{\Tcal}[M_{\Tcal}XM_{\Tcal}^{\mathsf T}]
 =\E_{\Tcal}[C_{\Tcal}XC_{\Tcal}^{\mathsf T}],
\]
where the last equality uses $JX=XJ=0$.
The operator $X\mapsto C_{\Tcal}XC_{\Tcal}^{\mathsf T}$ has trace $(\tr C_{\Tcal})^2$: its diagonal coefficients in the matrix-unit basis are $C_{\Tcal}(i,i)C_{\Tcal}(j,j)$. Since $C_{\Tcal}J=JC_{\Tcal}=0$, it vanishes on $\mathcal Z^\perp$, and its trace on $\mathcal Z$ is unchanged. Taking expectations and applying \cref{lem:three-copy} gives
\begin{equation}
 \E_{\Real}F^2=\Tr_{\mathcal Z}(\mathcal A^m).
 \label{eq:second-moment-trace}
\end{equation}
It remains to determine the spectrum of $\mathcal A$ on $\mathcal Z$.

\begin{samepage}
\begin{lemma}[Averaging-operator spectrum]
\label[lemma]{lem:spectrum}
On $\mathcal Z$, the operator $\mathcal A$ has the following eigenvalues and multiplicities:
\begin{center}
\begin{tabular}{@{}l >{$}r<{$}@{$\;=\;$}>{$}l<{$} c@{}}
\toprule
Mode & \multicolumn{2}{c}{Eigenvalue} & Multiplicity\\
\midrule
Common symmetric component & \lambda_1 & \lambda=1-1/n & $1$\\[2pt]
Symmetric vertex effects & \lambda_2 & 1-3/(2n) & $n-1$\\[2pt]
Symmetric residual edge effects & \lambda_3 & \lambda^2=(1-1/n)^2 & $n(n-3)/2$\\[2pt]
Skew-symmetric component & \lambda_4 & 1-2/n & $(n-1)(n-2)/2$\\
\bottomrule
\end{tabular}
\end{center}
\end{lemma}
\end{samepage}

\begin{proof}
Simultaneously relabeling rows and columns gives a representation $\rho$ of $\Sn$ on $\mathcal Z$:
\[
 \rho(\sigma)X=R_\sigma X R_\sigma^{\mathsf T}.
\]
Let $H=I-J$. The following four subspaces of $\mathcal Z$ are preserved by every relabeling, in the order listed in the lemma:
\[
\begin{array}{ll}
\text{common:}&\operatorname{span}\{H\},\\[2pt]
\text{vertex:}&\{H\operatorname{diag}(z)H:\ \sum_i z_i=0\},\\[2pt]
\text{residual:}&\{X\in\mathcal Z:X^{\mathsf T}=X,\ X_{ii}=0\text{ for all }i\},\\[2pt]
\text{skew:}&\{X\in\mathcal Z:X^{\mathsf T}=-X\}.
\end{array}
\]
Here $\operatorname{diag}(z)$ is the diagonal matrix with diagonal $z$. These subspaces have the dimensions in the lemma and form a direct sum: split into symmetric and skew-symmetric parts, then use the common and vertex components to match the diagonal, leaving the residual component. Indeed, $(H\operatorname{diag}(z)H)_{ii}=(1-2/n)z_i$ for zero-sum $z$.

To justify irreducibility, let $V=\one^\perp$ be the standard zero-sum-vector representation. The map $u\otimes v\mapsto uv^{\mathsf T}$ identifies its tensor square $V\otimes V$ with $\mathcal Z$ and respects relabeling, since $(R_\sigma u)(R_\sigma v)^{\mathsf T}=R_\sigma(uv^{\mathsf T})R_\sigma^{\mathsf T}$. For $n\ge4$, this tensor square has exactly four irreducible summands~\cite[Exercise~4.19]{FH91}. Our four nonzero invariant subspaces must therefore each be \emph{irreducible}: none contains a smaller nonzero subspace preserved by every relabeling.

Let the operator $\mathcal K$ be the average over transpositions of distinct labels:
\[
 \mathcal K=\frac1{\binom n2}\sum_{i<j}\rho((i\,j)).
\]
This operator preserves each component and commutes with every relabeling. It is self-adjoint for the Frobenius inner product, so its eigenspaces are also preserved by every relabeling. By Schur's lemma, $\mathcal K$ multiplies each irreducible component $W$ by a scalar. All transpositions have the same trace on $W$, giving that scalar as
\begin{equation}
 r_W=\frac{\Tr_W\rho((1\,2))}{\dim W}.
 \label{eq:trace-ratio}
\end{equation}
Thus only the trace of one transposition on each component is needed.

To relate this relabeling average to our shuffle, observe that
$P_\tau=(I+R_{\tau})/2$. With independent uniform endpoints,
\[
 \E_\tau R_{\tau}=\left(1-\frac2n\right)I+\frac2nJ,
 \qquad
 \E_\tau \rho(\tau)=\frac1nI_{\mathcal Z}+\frac{n-1}{n}\mathcal K.
\]
The identity term in the second formula accounts for equal endpoints. Since $JX=XJ=0$ for $X\in\mathcal Z$, expansion gives
\begin{align}
 \mathcal A(X)
 &=\frac14\E_\tau\bigl[X+R_{\tau}X+XR_{\tau}
                         +R_{\tau}XR_{\tau}^{\mathsf T}\bigr]\notag\\
 &=\frac{n-1}{4n}\bigl(3X+\mathcal K(X)\bigr).
 \label{eq:class-average}
\end{align}
Hence its eigenvalue on $W$ is $(n-1)(3+r_W)/(4n)$.

It remains to compute the four traces and the dimensions of the respective eigenspaces. On the common component the trace is $1$. On the vertex component, relabeling simply permutes $z$, so the trace is $n-3$: a transposition fixes $n-2$ coordinate vectors, and the constant direction contributes~$1$.

The symmetric matrices in $\mathcal Z$ have basis $J_{(i\,j)}=\tfrac12(\mathbf e_i-\mathbf e_j)(\mathbf e_i-\mathbf e_j)^{\mathsf T}$, $i<j$, which is permuted by relabeling. The transposition $(1\,2)$ fixes $\{1,2\}$ and the $\binom{n-2}{2}$ pairs avoiding those labels. Subtracting the common and vertex traces gives the residual trace
\[
 1+\binom{n-2}{2}-1-(n-3)=\frac{(n-3)(n-4)}2.
\]
On all skew-symmetric matrices, the trace is $\binom{n-2}{2}-1$: the basis matrix on $\{1,2\}$ changes sign, disjoint pairs are fixed, and the other basis matrices are exchanged. This space splits into the skew component of $\mathcal Z$ and the matrices $z\one^{\mathsf T}-\one z^{\mathsf T}$ with $\sum_i z_i=0$. The latter transform as zero-sum vectors and have trace $n-3$. The skew component therefore has trace
\[
 \binom{n-2}{2}-1-(n-3)=\frac{(n-2)(n-5)}2.
\]
Dividing these traces by the dimensions in \cref{eq:trace-ratio} yields, in order,
\[
 r_W=1,\qquad \frac{n-3}{n-1},\qquad
 \frac{n-4}{n},\qquad \frac{n-5}{n-1}.
\]
Substitution into \eqref{eq:class-average} gives the four stated eigenvalues; their multiplicities are the component dimensions.
\end{proof}

\begin{proposition}[Variance under the realized distribution]
\label[proposition]{prop:second-moment}
For $n\ge4$ and $m\ge0$,
\begin{equation}
 \Var_{\Real}F\le\lambda^m+(n-1)\lambda_2^m.
 \label{eq:real-variance}
\end{equation}
\end{proposition}

\begin{proof}
By \eqref{eq:second-moment-trace}, we sum the $m$th powers of the eigenvalues in \cref{lem:spectrum}, with their multiplicities:
\begin{align}
 \E_{\Real}F^2
 &=\lambda^m+(n-1)\lambda_2^m
  +\frac{n(n-3)}2\lambda^{2m}
  +\frac{(n-1)(n-2)}2\lambda_4^m
  \label{eq:exact-second}\\
  &\le\lambda^m+(n-1)\lambda_2^m
  +\bigl((n-1)^2-n\bigr)\lambda^{2m},\notag
\end{align}
where the last inequality follows from $0\le \lambda_4=1-2/n\le(1-1/n)^2=\lambda^2$. Recalling that $\E_{\Real}F=(n-1)\lambda^m$ (\cref{lem:first}), we have
\[
 \Var_{\Real}F
 \le\lambda^m+(n-1)\lambda_2^m-n\lambda^{2m}
 \le\lambda^m+(n-1)\lambda_2^m.
\]
\end{proof}

The variance of $F$ under $\Real$ can also be bounded without determining
the full operator spectrum. One approach is to write $F$ as a sum of $n$
random variables using \cref{eq:statistic} and compute their pairwise
covariances. For $m=2n\log n+O(n)$, the total covariance contribution is
lower order than the sum of the individual variances. We use the spectral
argument because, although less elementary, it makes the structure of the
second moment more transparent.

\subsection{The lower bound and the cutoff window}

The moment estimates now give the lower bound by Chebyshev's inequality.

\begin{theorem}[Explicit lower bound]
\label[theorem]{thm:lower}
For $n\ge4$ and $m\ge0$, with $\lambda$ and $\lambda_2$ as in \cref{lem:spectrum},
\begin{equation}
 d_n(m)\ge
 \pos{1-\frac8{(n-1)^2\lambda^m}
       -\frac4{n-1}\left(\frac{\lambda_2}{\lambda^2}\right)^m}.
 \label{eq:lower}
\end{equation}
\end{theorem}

\begin{proof}
Write $s=(n-1)\lambda^m>0$ for the mean under the realized distribution. The ideal mean is zero, so we distinguish the two experiments using $\{F\ge s/2\}$. Chebyshev's inequality, \cref{lem:first}, and \cref{prop:second-moment} give
\begin{align*}
 d_n(m)
 &\ge1-\Pp_{\Real}(F<s/2)-\Pp_{\Ideal}(F\ge s/2)\\
 &\ge1-\frac{4(\lambda^m+(n-1)\lambda_2^m)}{(n-1)^2\lambda^{2m}}
          -\frac{4\lambda^m}{(n-1)^2\lambda^{2m}}\\
 &=1-\frac8{(n-1)^2\lambda^m}
       -\frac4{n-1}\left(\frac{\lambda_2}{\lambda^2}\right)^m.
\end{align*}
Taking the positive part is valid because TV distance is nonnegative.

\end{proof}

\begin{proof}[Proof of \cref{thm:main}]
Write $c=m/n-2\log n$. Suppose first that $c<0$, and put
$\rho_m=((n-1)^2\lambda^m)^{-1}$.
Convexity gives $\lambda_2\le\lambda^{3/2}$, so \cref{thm:lower} yields
\[
 1-d_n(m)\le8\rho_m+4\sqrt{\rho_m}.
\]
Using $-\log\lambda\le1/n+1/[2n(n-1)]$ and
$\log(n/(n-1))\le1/(n-1)$, we obtain
\[
 \log\rho_m\le c+\frac{\log n+2}{n-1}\le c+\log2
 \qquad(n\ge10).
\]
Set $z=e^{c/2}$. If $z\ge1/8$, \eqref{eq:left-window} is trivial.
Otherwise,
\[
 1-d_n(m)\le16z^2+4\sqrt2\,z
 \le(2+4\sqrt2)z<8z,
\]
proving the left-hand bound.

If $c\ge4$, \cref{prop:upper} gives
$d_n(m)\le(e^2/2)e^{-c/2}\le8e^{-c/2}$.
For $0<c<4$, the same bound follows from $d_n(m)\le1<8e^{-2}$.
This proves \eqref{eq:right-window}.
Finally, monotonicity of $d_n$ and these two bounds place each fixed-error
mixing time within $O_\eps(n)$ of $2n\log n$, proving \eqref{eq:mixing-time}.
\end{proof}

The test function can be computed from the schedule and the candidate permutation. Starting from the identity matrix, form $M_{\Tcal}$ by averaging the corresponding rows, and then evaluate \eqref{eq:statistic}, aggregating single-card biases across the entire deck.

\section{A shallow public switching network}
\label{sec:depth}

We now apply the mixing bound to parallel switching networks. Sample a schedule of a Monte shuffle. Form a directed dependency graph where two switches are connected if they share an endpoint; the arc goes in the direction of their order in the schedule. Switches that are not logically dependent can be executed in parallel.

Write $H_m$ for the number of parallel layers, equivalently the maximum number of switches on a directed dependency path. This is a random variable determined by the public schedule. Czumaj and
Lingas~\cite{CL23} show that for
$m=\Omega(n\log n)$ independent uniform distinct-pair interactions,
the depth is $O(m/n)$ in expectation and with high probability.
Their path-counting estimate supplies the constants below. It also
applies to our with-replacement schedule: replacing every public null
step by a fresh uniform distinct pair gives their model, and deleting
those replacements cannot increase depth.

\begin{corollary}[A shallow public switching network]
\label[corollary]{cor:network}
Let $n\ge10$, $0<\eps\le1/2$, and $0<\delta<1$. Sampling a public
schedule of length
\begin{equation}
 m=\ceil{2n\left(\log n+2+\log\frac1{2\eps}
                 +\frac12\log\frac2\delta\right)}
 \label{eq:network-size}
\end{equation}
and parallelizing its nonnull switches yields, with probability at
least $1-\delta$, a network with conditional TV error at most $\eps$,
at most $m$ switches, and depth at most
\begin{equation}
 \ceil{\max\left\{\frac{16em}{n},\,\log_2\frac{2n}{\delta}\right\}}.
 \label{eq:network-depth}
\end{equation}
For fixed $\eps,\delta$, these bounds are
$2n\log n+O_{\eps,\delta}(n)$ switches and
$32e\log n+O_{\eps,\delta}(1)$ layers.
\end{corollary}

\begin{proof}
Equation~\eqref{eq:eps-delta-upper} bounds the mixing failure
probability by $\delta/2$. To complete the union bound, let $k$ be the
ceiling in \eqref{eq:network-depth}. The path-counting estimate of
Czumaj and Lingas~\cite[Section~4]{CL23}, with $k!\ge(k/e)^k$, gives
\[
 \Pp(H_m\ge k)
 \le\frac{n(8m/n)^k}{k!}
 \le n\left(\frac{8em}{nk}\right)^k
 \le n2^{-k}\le\frac\delta2.\qedhere
\]
\end{proof}

\clearpage
\begingroup
\small
\raggedright
\AddToHookNext{env/thebibliography/begin}{%
  \setlength{\itemsep}{0pt}%
  \setlength{\parskip}{0pt}%
}
\bibliographystyle{alphaurl}
\bibliography{references}
\endgroup
\end{document}